\documentclass{iopjournal}

\usepackage[T1]{fontenc}
\usepackage[utf8]{inputenc}
\DeclareUnicodeCharacter{FFFC}{}

\usepackage{mathrsfs}
\usepackage{amsfonts}
\usepackage{epsfig}
\usepackage{amsthm}
\usepackage{amsmath}
\usepackage{amssymb}
\usepackage{graphicx}
\usepackage{verbatim}
\usepackage{booktabs}
\usepackage{subcaption}
\usepackage{array,multirow}
\usepackage{url}
\usepackage{color}
\usepackage{cprotect}
\usepackage{tabularx}
\usepackage{dcolumn}
\usepackage{bm}
\usepackage{slashed}
\usepackage{physics}
\usepackage{framed}
\usepackage{orcidlink}
\usepackage{enumitem}

\definecolor{nicered}{rgb}{0.5,0.,0.}
\definecolor{nicegreen}{rgb}{0.,0.5,0.}
\definecolor{niceblue}{rgb}{0.,0.,0.5}

\setlist{nolistsep}

\newcommand{\calR}{\mathcal R} 
\newcommand{\calA}{\mathcal A} 
\newcommand{\calW}{\mathcal W} 
\newcommand{\calC}{\mathcal C} 
\newcommand{\calM}{\mathcal M} 
\newcommand{\calI}{\mathcal I} 
 
\newcommand{\TTNC}{\mathrm{TTNC}} 
\newcommand{\ren}{\mathrm{ren}} 
\newcommand{\reg}{\mathrm{reg}} 
\newcommand{\ct}{\mathrm{ct}} 
\newcommand{\loc}{\mathrm{loc}}

\newcommand{\AdS}{\mathrm{AdS}}

\begin{document}

\articletype{Paper}

\title{Matter one-loop logarithms and homogeneous TTNC scale response of Lifshitz black branes}

\author{Yingnan Xu$^{1,2,*}$\orcid{0009-0005-8296-0850} and Shuangshuang Chu$^3$\orcid{0009-0004-4396-582X}}

\affil{$^1$Zhongtai Securities Institute for Financial Studies, Shandong University,
Jinan, Shandong 250014, China}

\affil{$^2$Department of Physics, Southern Methodist University,
Dallas, Texas 75206, U.S.A.}

\affil{$^3$School of Statistics, Dongbei University of Finance and Economics,
Dalian, Liaoning 116025, China}

\affil{$^*$Author to whom any correspondence should be addressed.}

\email{yingnanx@mail.smu.edu}


\begin{abstract}
We compute the logarithmic one-loop matter contribution to the thermodynamics
and homogeneous twistless torsional Newton--Cartan scale response of a
four-dimensional Lifshitz black brane.  The background is the neutral planar
member of the analytic Einstein--Maxwell--dilaton Lifshitz black-brane family,
while the quantum fields are treated as probes: a real scalar with arbitrary
mass and nonminimal curvature coupling, a four-component Dirac spinor, and an
Abelian Maxwell field with its Faddeev--Popov ghosts.  For each spin sector,
the logarithmic coefficient separates into a smooth radial heat-kernel
contribution, governed by \(\calC_1\), and a horizon-localized conical
contribution, governed by \(\calW_h\).  This separation identifies which part
of the matter one-loop logarithm is visible in the boundary Lifshitz/TTNC Ward
identity and which part instead belongs to horizon replica entropy.  The smooth coefficient \(\calC_1\) controls the source-induced
homogeneous projection of the TTNC Weyl Ward identity, namely the smooth
boundary-source contribution to the Ward combination \(z\epsilon-2p\),
whereas the thermodynamic logarithmic entropy is controlled by
\(\calC_{\rm therm}=\calC_1+zL^2\calW_h\).  We give closed expressions for
\(\calC_1\), \(\calW_h\), and \(\calC_{\rm therm}\) for the scalar, Dirac,
and Maxwell probe sectors, identify the gauge-field contact contribution in
the conical entropy, and verify that the smooth source response vanishes in
the relativistic planar \(z=1\) limit while the standard horizon heat-kernel
entropy coefficient remains.
\end{abstract}

\section{Introduction}
\label{sec:introduction}

Quantum field theories with anisotropic Lifshitz scaling provide a useful
arena for studying strongly coupled systems without relativistic conformal
symmetry.  Gauge/gravity duality realizes this scaling geometrically through
asymptotically Lifshitz spacetimes
\cite{Kachru:2008yh,Taylor:2008tg,Taylor:2015glc}.  At finite temperature,
Lifshitz black holes and black branes provide the corresponding thermal
saddles and exhibit characteristic scaling relations among temperature,
entropy, energy, and pressure
\cite{Mann:2009yx,Bertoldi:2009vn,Bertoldi:2009dt,Tarrio:2011de,
Ross:2011gu,Cong:2024lfh}.

A central feature of Lifshitz holography is the nonrelativistic geometric
structure induced at the boundary.  In the vielbein formulation, the boundary
sources organize into torsional Newton--Cartan data, with the twistless
sector described by TTNC geometry
\cite{Christensen:2013rfa,Christensen:2013lma,Bergshoeff:2014uea,
Hartong:2015zia,Hartong:2022lsy}.  The associated anisotropic Weyl Ward
identity relates the energy density, the spatial stress trace, and the
Lifshitz/TTNC scale anomaly.  For homogeneous thermal states this Ward
identity gives a direct and invariant way to identify the scale dependence of
the renormalized generating functional
\cite{Baggio:2011ha,Arav:2014goa,Arav:2016xjc,Auzzi:2015fgg,
Auzzi:2016lrq,Arav:2016akx}.

The physical question addressed here is how quantum matter modifies the
homogeneous Lifshitz scale relation.  For flat boundary sources, the
classical black brane satisfies the scale equation of state
\(z\epsilon=2p\).  At one loop, logarithmic terms introduce
renormalization-scale dependence into the thermal generating functional and
can therefore generate a quantum contribution to the Ward combination
\(z\epsilon-2p\).  To identify this contribution, one must distinguish the
part of the logarithmic scale dependence associated with the boundary TTNC
sources from the part localized at the horizon as a replica-entropy
contribution.  We study this question for the four-dimensional neutral
planar Lifshitz black brane of Tarrio and Vandoren
\cite{Tarrio:2011de}. This background is especially useful because it is an
analytic finite-temperature Lifshitz saddle with homogeneous flat TTNC
boundary sources, so the source-side scale response can be isolated without
additional complications from spatial curvature, inhomogeneity, or
background charge. We keep the background fixed and treat the quantum
fields as probes.  The probe sectors consist of a real scalar with arbitrary
mass and nonminimal curvature coupling, a four-component Dirac spinor, and
an Abelian Maxwell field together with its gauge-fixed ghost determinant.
The scalar sector extends curved-space QFT analyses of fields in Lifshitz
black-hole backgrounds, including the vacuum-polarization study of Quinta,
Flachi, and Lemos \cite{Quinta:2016cvy}, while the spinor and vector sectors
complete the corresponding spin-dependent probe analysis and provide a
useful reference for future calculations of the full coupled
Einstein--Maxwell--dilaton fluctuation determinant.

The main result is that the logarithmic sector separates into two local
contributions with distinct geometric and physical meanings.  The first is a
smooth radial logarithm in the integrated bulk heat-kernel coefficient.  Its
dimensionless coefficient will be denoted by
\(\calC_{\rm sm}\equiv\calC_1\).  This term determines the logarithmic
counterterm for the boundary TTNC sources and therefore controls the smooth
source-induced contribution to the anisotropic Weyl Ward identity.  In the
homogeneous thermal state, it is the coefficient that contributes to the
quantum correction of \(z\epsilon-2p\).  The second contribution is a
horizon-localized conical term in the replica entropy.  Its thermodynamic
coefficient is \(\calC_{\Sigma}\equiv zL^2\calW_h\), where \(\calW_h\) is
the local conical density on the horizon.  The total logarithmic thermal
coefficient is therefore
\begin{equation} 
\calC_{\rm therm} = \calC_{\rm sm}+\calC_{\Sigma} = \calC_1+zL^2\calW_h. 
\label{eq:Ctherm-def-intro} 
\end{equation}
Thus the same one-loop heat-kernel calculation distinguishes the logarithmic
response of the boundary sources from the logarithmic correction to the
thermal entropy.  The distinction is essential because the smooth TTNC Ward
projection is governed by \(\calC_1\), whereas the logarithmic thermal
entropy is governed by the combined coefficient
\(\calC_{\rm therm}\).  The split therefore provides a direct way to track
how ultraviolet source renormalization and horizon replica physics enter
Lifshitz black-brane thermodynamics.

The resulting coefficients give a compact spin-dependent characterization of
matter one-loop effects on the Lifshitz saddle.  They also provide a useful
check on the relativistic planar limit.  For \(z=1\), the smooth radial
coefficient vanishes in the flat-boundary source sector, while the conical
horizon coefficient remains and reproduces the standard local heat-kernel
entropy contribution.  Thus the calculation isolates an intrinsically
anisotropic quantum scale response, encoded in \(\calC_1\), from the ordinary
horizon entropy logarithm, encoded in \(\calW_h\).  In this sense it supplies
both a consistency test of the Lifshitz/TTNC renormalization structure and a
reference point for the more involved dynamical EMD one-loop problem.

The paper is organized as follows.  In section~\ref{sec:background} we
review the Lifshitz black-brane background and the associated homogeneous
TTNC boundary data.  In section~\ref{sec:matter-determinants} we define the
scalar, Dirac, and Maxwell probe determinants and the heat-kernel
coefficients used in the calculation.  In section~\ref{sec:horizon-entropy}
we compute the conical horizon densities.  In section~\ref{sec:smooth-log}
we extract the smooth radial logarithms.  In section~\ref{sec:coefficients}
we collect the spin-dependent coefficients and their total matter
combination.  In section~\ref{sec:holo-ren} we relate the smooth radial
logarithm to the TTNC logarithmic counterterm, reconstruct the logarithmic
thermodynamics, identify the homogeneous Ward projection, and state the
source-side heat-kernel representation relevant for non-flat TTNC sources.
In section~\ref{sec:z1-check} we discuss the relativistic planar \(z=1\)
limit.  Section~\ref{sec:discussion} contains the discussion and concludes the paper.
Appendices~\ref{app:cone} and \ref{app:curvature-expansion} contain the
conical and radial heat-kernel derivation details.

\section{The Lifshitz black-brane background}
\label{sec:background}

We work in four bulk dimensions and take the boundary theory to have two spatial directions.  As a concrete exact saddle we use the neutral planar member of the Einstein--Maxwell--dilaton Lifshitz black-brane family of Tarrio and Vandoren~\cite{Tarrio:2011de}.  The bulk theory has the schematic form
\begin{equation}
I_{\rm EMD}
=
\frac{1}{16\pi G_4}
\int \dd^4x \sqrt{-g}
\left[
\calR-2\Lambda
-\frac12(\partial\varphi)^2
-\frac14\sum_I e^{\lambda_I\varphi}F_I^2
\right].
\label{eq:EMD-action}
\end{equation}
Only the metric is needed below, since all quantum fields considered in this paper are probes.  In Lorentzian signature the neutral planar black brane is
\begin{equation}
\dd s^2
=
L^2\left[
-r^{2z}f(r)\dd t^2
+r^2(\dd x^2+\dd y^2)
+\frac{\dd r^2}{r^2f(r)}
\right],
\qquad
f(r)=1-\left(\frac{r_h}{r}\right)^{z+2}.
\label{eq:lif-bb}
\end{equation}
The Euclidean metric is obtained by \(t=-i\tau\),
\begin{equation}
\dd s_E^2
=
L^2\left[
r^{2z}f(r)\dd\tau^2
+r^2(\dd x^2+\dd y^2)
+\frac{\dd r^2}{r^2f(r)}
\right].
\label{eq:euclidean-lif-bb}
\end{equation}
Regularity of the Euclidean cigar at \(r=r_h\) fixes the inverse temperature,
\begin{equation}
T=\frac{1}{\beta}=\frac{z+2}{4\pi}r_h^z .
\label{eq:temperature}
\end{equation}
The horizon area density and classical entropy density are
\begin{equation}
\frac{A_h}{V_2}=L^2r_h^2,
\qquad
s_0=\frac{L^2r_h^2}{4G_4}.
\label{eq:classical-entropy}
\end{equation}

Near the boundary, the induced Lifshitz geometry is naturally described in terms of torsional Newton--Cartan data~\cite{Christensen:2013rfa,Christensen:2013lma}.  For the homogeneous black brane these sources reduce to flat TTNC data,
\begin{equation}
\tau=\dd t,
\qquad
h_{ij}\dd x^i\dd x^j=\dd x^2+\dd y^2,
\qquad
\tau\wedge\dd\tau=0 .
\label{eq:flat-ttnc}
\end{equation}
The corresponding anisotropic Weyl transformation acts as
\begin{equation}
\tau_\mu\to e^{z\sigma}\tau_\mu,
\qquad
h_{\mu\nu}\to e^{2\sigma}h_{\mu\nu}.
\label{eq:lif-weyl}
\end{equation}
For a homogeneous equilibrium state in \(2+1\) boundary dimensions the thermal projection of the Lifshitz/TTNC Weyl Ward identity takes the form
\begin{equation}
z\varepsilon-2p=\calA_{\TTNC}.
\label{eq:ttnc-ward}
\end{equation}
Classically, for flat boundary sources, \(\calA_{\TTNC}=0\) and therefore \(z\varepsilon_0=2p_0\).

\section{Probe matter determinants and heat-kernel coefficients}
\label{sec:matter-determinants}

We now define the probe one-loop determinants.  The fields considered here do not represent the coupled graviton--gauge--dilaton fluctuation system of the background EMD theory.  Rather, they are probe matter sectors on the fixed geometry \eqref{eq:euclidean-lif-bb}.  This distinction is important for the Maxwell field: the vector determinant below is an Abelian probe determinant and includes its Faddeev--Popov ghosts, but it is not the determinant of the background-supporting EMD gauge fluctuations mixed with the metric and dilaton.

\subsection{Real scalar}
\label{subsec:scalar-det}

The scalar action is
\begin{equation}
I_\chi
=
\frac12\int\dd^4x\sqrt g\,
\left[
g^{MN}\partial_M\chi\partial_N\chi
+m_\chi^2\chi^2
+\xi\calR\chi^2
\right].
\label{eq:scalar-action}
\end{equation}
After integrating by parts,
\begin{equation}
I_\chi
=
\frac12\int\dd^4x\sqrt g\,\chi\Delta_\chi\chi,
\qquad
\Delta_\chi=-\nabla^2+m_\chi^2+\xi\calR .
\label{eq:scalar-operator}
\end{equation}
The one-loop effective action is
\begin{equation}
\Gamma_\chi=\frac12\log\det\frac{\Delta_\chi}{\mu^2}.
\label{eq:scalar-oneloop}
\end{equation}
The relevant four-dimensional heat-kernel expansion is
\begin{equation}
\Tr e^{-s\Delta}
\sim
\frac{1}{(4\pi s)^2}
\int\dd^4x\sqrt g\,
\left[a_0+a_2s+a_4s^2+\cdots\right].
\label{eq:heat-kernel-expansion}
\end{equation}
For the scalar operator \eqref{eq:scalar-operator}, and dropping total derivatives, the coefficient is \cite{Vassilevich:2003xt}
\begin{equation}
a_4^{(\chi)}
=
\frac{1}{180}\left(
\calR_{MNRS}\calR^{MNRS}-\calR_{MN}\calR^{MN}
\right)
+\frac12\nu_\xi^2\calR^2
+\nu_\xi m_\chi^2\calR
+\frac12m_\chi^4,
\label{eq:a4-scalar}
\end{equation}
where
\begin{equation}
\nu_\xi\equiv \xi-\frac16 .
\label{eq:nuxi}
\end{equation}

\subsection{Dirac spinor}
\label{subsec:spinor-det}

For one four-component Euclidean Dirac spinor,
\begin{equation}
\Gamma_{\psi}
=
-\log\det(\slashed\nabla+m_\psi)
=
-\frac12\log\det\frac{\Delta_{1/2}}{\mu^2},
\qquad
\Delta_{1/2}=-\nabla^2+\frac14\calR+m_\psi^2 .
\label{eq:spinor-det}
\end{equation}
The minus sign is the Grassmann sign.  Including this sign in the local density, the relevant coefficient is
\begin{equation}
\mathfrak a_4^{(\psi)}
=
\frac{7}{360}\calR_{MNRS}\calR^{MNRS}
+\frac{1}{45}\calR_{MN}\calR^{MN}
-\frac{1}{72}\calR^2
-\frac13m_\psi^2\calR
-2m_\psi^4 .
\label{eq:a4-spinor}
\end{equation}
The notation \(\mathfrak a_4\) indicates that the statistics sign has already been included.

\subsection{Maxwell vector}
\label{subsec:vector-det}

For one Abelian Maxwell probe in Feynman gauge,
\begin{equation}
\Gamma_A
=
\frac12\log\det\frac{\Delta_1}{\mu^2}
-
\log\det\frac{\Delta_0}{\mu^2},
\label{eq:vector-det}
\end{equation}
where
\begin{equation}
(\Delta_1)_M{}^N
=
-\delta_M{}^N\nabla^2+\calR_M{}^N,
\qquad
\Delta_0=-\nabla^2 .
\label{eq:vector-operators}
\end{equation}
The second determinant is the complex Faddeev--Popov ghost determinant.  The signed gauge-plus-ghost heat-kernel density is
\begin{equation}
\mathfrak a_4^{(A)}
=
-\frac{13}{180}\calR_{MNRS}\calR^{MNRS}
+\frac{88}{180}\calR_{MN}\calR^{MN}
-\frac{25}{180}\calR^2 .
\label{eq:a4-vector}
\end{equation}
The conical entropy extracted from this gauge-fixed determinant contains the standard spin-one contact contribution at the horizon.  This is the appropriate local coefficient for the gauge-fixed one-loop gravitational entropy of the probe Maxwell field~\cite{Kabat:1995eq}.

\section{Conical horizon densities}
\label{sec:horizon-entropy}

Let \(\Sigma\) denote the horizon cross-section.  For a stationary horizon, the logarithmic entropy associated with the local one-loop effective action can be obtained from the replicated geometry \(\calM_n\) with opening angle \(2\pi n\) around \(\Sigma\) \cite{Fursaev:1995ef,Solodukhin:2011gn}.  We introduce the normal-plane curvature contractions
\begin{equation}
\calR_{\perp}=\calR_{\hat\tau\hat\tau}+\calR_{\hat r\hat r},
\qquad
\calR_{\perp\perp}=\calR_{\hat\tau\hat r\hat\tau\hat r},
\label{eq:normal-curvatures}
\end{equation}
where hatted indices denote an orthonormal frame in the Euclidean \((\tau,r)\) plane.

For any signed local density of the form
\begin{equation}
\mathfrak a_4
=
\alpha\calR_{MNRS}\calR^{MNRS}
+\beta\calR_{MN}\calR^{MN}
+\gamma\calR^2
+\delta m^2\calR+\text{terms without curvature},
\label{eq:generic-a4}
\end{equation}
the corresponding local conical density is
\begin{equation}
\calW_h
=
2\alpha\calR_{\perp\perp,h}
+\beta\calR_{\perp,h}
+2\gamma\calR_h
+\delta m^2 .
\label{eq:generic-Wh}
\end{equation}
For the Lifshitz black brane, with curvature convention \([\nabla_M,\nabla_N]V^P=\calR_{MNQ}{}^PV^Q\), the required horizon curvatures are
\begin{align}
\calR_h&=-\frac{2(z+1)(z+2)}{L^2},
\label{eq:Rh-horizon}\\
\calR_{\perp,h}&=-\frac{2z(z+2)}{L^2},
\label{eq:Rperp-horizon}\\
\calR_{\perp\perp,h}&=-\frac{(z-1)(z+2)}{L^2}.
\label{eq:Rperpperp-horizon}
\end{align}

The scalar density \eqref{eq:a4-scalar} gives
\begin{equation}
\calW_h^{(\chi)}
=
\nu_\xi m_\chi^2
-\frac{2(z+1)(z+2)}{L^2}\nu_\xi^2
+\frac{z+2}{90L^2}.
\label{eq:Wh-scalar}
\end{equation}
The Dirac density \eqref{eq:a4-spinor} gives
\begin{equation}
L^2\calW_h^{(\psi)}
=
\frac{34+7z-5z^2-60(m_\psi L)^2}{180}.
\label{eq:Wh-spinor}
\end{equation}
The Maxwell gauge-plus-ghost density \eqref{eq:a4-vector} gives
\begin{equation}
L^2\calW_h^{(A)}
=
-\frac{(z+2)(25z-37)}{90}.
\label{eq:Wh-vector}
\end{equation}
The corresponding horizon-localized logarithmic entropy density for any one of the probe sectors is
\begin{equation}
s_{1,\log}^{\Sigma,(s)}
=
\frac{L^2}{4\pi}\calW_h^{(s)}r_h^2\log\frac{r_h}{\mu},
\label{eq:slog-horizon-general}
\end{equation}
where \(s=\chi,\psi,A\).  For the Maxwell field this expression includes the gauge-field contact contribution associated with the gauge-fixed determinant.

\section{Smooth radial logarithms}
\label{sec:smooth-log}

The horizon coefficient \(\calW_h\) is the localized conical contribution.  The smooth part of the one-loop determinant contains a logarithmic radial term in the integrated heat-kernel coefficient.  Let
\begin{equation}
u(r)\equiv \left(\frac{r_h}{r}\right)^{z+2} .
\label{eq:u-def}
\end{equation}
For any signed density \(\mathfrak a_4^{(s)}\), the large-\(r\) expansion takes the form
\begin{equation}
L^4\mathfrak a_4^{(s)}(r)=A_0^{(s)}+\calC_1^{(s)}u(r)+O(u^2),
\label{eq:a4-large-r-general}
\end{equation}
where \(A_0^{(s)}\) is independent of \(r_h\).  Since
\begin{equation}
\sqrt g=L^4r^{z+1},
\label{eq:sqrtg}
\end{equation}
the term linear in \(u\) gives
\begin{equation}
\int_{r_h}^{\Lambda}\dd r\sqrt g\,\mathfrak a_4^{(s)}
\supset
\beta V_2\calC_1^{(s)}r_h^{z+2}\int_{r_h}^{\Lambda}\frac{\dd r}{r}
=
\beta V_2\calC_1^{(s)}r_h^{z+2}\log\frac{\Lambda}{r_h}.
\label{eq:radial-log-general}
\end{equation}

The required curvature expansions are collected in appendix~\ref{app:curvature-expansion}.  Substituting them into \eqref{eq:a4-scalar}, \eqref{eq:a4-spinor}, and \eqref{eq:a4-vector} gives
\begin{align}
\calC_1^{(\chi)}
&=
\frac{z(z-1)(z-4)}{45}
+4(z-1)(z^2+2z+3)\nu_\xi^2
-2(z-1)(m_\chi L)^2\nu_\xi,
\label{eq:C1-scalar}\\
\calC_1^{(\psi)}
&=
\frac{2(z-1)}{15}\left[5(m_\psi L)^2+z(z-4)\right],
\label{eq:C1-spinor}\\
\calC_1^{(A)}
&=
\frac{4z(z-1)(z-4)}{15}.
\label{eq:C1-vector}
\end{align}
All three smooth radial coefficients vanish identically at \(z=1\).  Thus the smooth source-induced logarithm is intrinsically anisotropic for this planar black-brane family.

\section{Spin-dependent coefficients and matter sums}
\label{sec:coefficients}

For each probe sector we define
\begin{equation}
\calC_{\rm sm}^{(s)}\equiv\calC_1^{(s)},
\qquad
\calC_{\Sigma}^{(s)}\equiv zL^2\calW_h^{(s)},
\qquad
\calC_{\rm therm}^{(s)}\equiv \calC_1^{(s)}+zL^2\calW_h^{(s)} .
\label{eq:sector-coeff-defs}
\end{equation}
Using \eqref{eq:Wh-scalar}--\eqref{eq:Wh-vector} and \eqref{eq:C1-scalar}--\eqref{eq:C1-vector}, the total thermal coefficients are
\begin{align}
\calC_{\rm therm}^{(\chi)}
&=
(2-z)(m_\chi L)^2\nu_\xi
+2(z^3-z^2-6)\nu_\xi^2
+\frac{z(2z^2-9z+10)}{90},
\label{eq:Ctherm-scalar}\\
\calC_{\rm therm}^{(\psi)}
&=
\frac{60(z-2)(m_\psi L)^2+z(19z^2-113z+130)}{180},
\label{eq:Ctherm-spinor}\\
\calC_{\rm therm}^{(A)}
&=
-\frac{z(z^2+133z-170)}{90}.
\label{eq:Ctherm-vector}
\end{align}
The spin-dependent result is summarized in table~\ref{tab:coefficients}.

\begin{table}[t]
\centering
\scriptsize
\renewcommand{\arraystretch}{1.7}
\begin{tabularx}{\textwidth}{c>{\centering\arraybackslash}X>{\centering\arraybackslash}X>{\centering\arraybackslash}X}
\toprule
field & \(\calC_1\) & \(L^2\calW_h\) & \(\calC_{\rm therm}=\calC_1+zL^2\calW_h\) \\
\midrule
real scalar
&
\(\displaystyle
\frac{z(z-1)(z-4)}{45}
+4(z-1)(z^2+2z+3)\nu_\xi^2
-2(z-1)(m_\chi L)^2\nu_\xi
\)
&
\(\displaystyle
(m_\chi L)^2\nu_\xi
-2(z+1)(z+2)\nu_\xi^2
+\frac{z+2}{90}
\)
&
\(\displaystyle
(2-z)(m_\chi L)^2\nu_\xi
+2(z^3-z^2-6)\nu_\xi^2
+\frac{z(2z^2-9z+10)}{90}
\)
\\
Dirac spinor
&
\(\displaystyle
\frac{2(z-1)}{15}\left[5(m_\psi L)^2+z(z-4)\right]
\)
&
\(\displaystyle
\frac{34+7z-5z^2-60(m_\psi L)^2}{180}
\)
&
\(\displaystyle
\frac{60(z-2)(m_\psi L)^2+z(19z^2-113z+130)}{180}
\)
\\
Maxwell vector
&
\(\displaystyle
\frac{4z(z-1)(z-4)}{15}
\)
&
\(\displaystyle
-\frac{(z+2)(25z-37)}{90}
\)
&
\(\displaystyle
-\frac{z(z^2+133z-170)}{90}
\)
\\
\bottomrule
\end{tabularx}
\caption{Smooth radial, conical horizon, and total logarithmic thermal coefficients for probe fields on the four-dimensional Lifshitz black brane.  The Dirac and Maxwell entries include the fermionic statistics sign and the Maxwell Faddeev--Popov ghosts, respectively.}
\label{tab:coefficients}
\end{table}

For \(N_\chi\) real scalars, \(N_\psi\) four-component Dirac spinors, and \(N_A\) Maxwell probe fields, the matter coefficients are
\begin{align}
\calC_1^{\rm matter}
&=N_\chi\calC_1^{(\chi)}+N_\psi\calC_1^{(\psi)}+N_A\calC_1^{(A)},
\label{eq:C1-matter}\\
\calW_h^{\rm matter}
&=N_\chi\calW_h^{(\chi)}+N_\psi\calW_h^{(\psi)}+N_A\calW_h^{(A)},
\label{eq:Wh-matter}\\
\calC_{\rm therm}^{\rm matter}
&=\calC_1^{\rm matter}+zL^2\calW_h^{\rm matter}.
\label{eq:Ctherm-matter}
\end{align}
These formulae make explicit which part of the matter logarithm is a smooth source response and which part is a horizon replica contribution.

\section{Logarithmic renormalization, thermodynamics, and TTNC Ward projection}
\label{sec:holo-ren}

\subsection{Smooth radial logarithm and TTNC counterterm}
\label{subsec:smooth-log-counterterm}

We regulate the radial coordinate at \(r=\Lambda\).  The coefficient
\(\calC_1^{(s)}\) has a direct asymptotic origin.  Near the Lifshitz
boundary, the blackening factor enters through
\begin{equation}
  u(r)=\left(\frac{r_h}{r}\right)^{z+2},
\end{equation}
and the local heat-kernel density has the large-\(r\) expansion
\begin{equation}
  L^4 a_4^{(s)}(r)
  =
  A_0^{(s)}
  +
  \calC_1^{(s)}u(r)
  +
  O(u^2).
  \label{eq:a4-asymptotic-C1}
\end{equation}
Since \(\sqrt g=L^4r^{z+1}\), the term proportional to
\(\calC_1^{(s)}\) gives
\begin{equation}
  \sqrt{g}\,a^{(s)}_4\,d^4x
  \supset
  C^{(s)}_1 r_h^{z+2}\frac{dr}{r}\,d\tau\,d^2x .
  \label{eq:radial_log_density_form}
\end{equation}
Thus the coefficient linear in \(u\) is precisely the coefficient of the
homogeneous radial logarithm.  The leading term \(A_0^{(s)}\) gives the
zero-temperature Lifshitz power divergence, while the terms \(O(u^2)\) do
not produce the same homogeneous \(r_h^{z+2}\log\Lambda\) contribution.

For any probe sector, the smooth logarithmic part of the regulated one-loop
functional follows from \eqref{eq:radial-log-general}:
\begin{equation}
  \Gamma_{1,\reg}^{\rm sm,(s)}
  \supset
  \frac{\beta V_2}{16\pi^2}
  \calC_1^{(s)}r_h^{z+2}
  \log\frac{\Lambda}{r_h}.
  \label{eq:smooth-reg-log-general}
\end{equation}
This logarithmic divergence is removed by a boundary counterterm written in
terms of the induced TTNC sources,
\begin{equation}
  \Gamma_{1,\ct}^{\rm sm,(s)}
  =
  -\frac{\log(\Lambda/\mu)}{16\pi^2}
  \int_{\partial\calM_\Lambda}\dd^3x\,N\sqrt h\,
  \mathfrak A_{\TTNC,\rm sm}^{(s)} .
  \label{eq:smooth-counterterm-general}
\end{equation}
Here \(\mathfrak A_{\TTNC,\rm sm}^{(s)}\) denotes the smooth source-side
TTNC density associated with the probe operator. In the present homogeneous calculation the full off-shell TTNC density is
not required.  The radial heat-kernel computation fixes only its
homogeneous black-brane projection, i.e. the coefficient of the
temperature-dependent \(dr/r\) logarithm.  Evaluated on the Lifshitz
black-brane saddle, this projection is
\begin{equation}
  \left.
  \int_{\partial\calM_\Lambda}\dd^3x\,N\sqrt h\,
  \mathfrak A_{\TTNC,\rm sm}^{(s)}
  \right|_{\rm hom.\,BB}
  =
  \beta V_2\calC_1^{(s)}r_h^{z+2}.
  \label{eq:homogeneous-ATTNC-sector}
\end{equation}
This equation should be understood as the saddle-point value of the smooth logarithmic projection, not as the definition of a counterterm chosen with explicit dependence on the horizon radius. Combining \eqref{eq:smooth-reg-log-general} and
\eqref{eq:smooth-counterterm-general} gives the finite smooth logarithmic
free-energy density
\begin{equation}
  f_{1,\log}^{\rm sm,(s)}
  =
  -\frac{\calC_1^{(s)}}{16\pi^2}
  r_h^{z+2}\log\frac{r_h}{\mu}.
  \label{eq:flog-smooth-sector}
\end{equation}

Under an infinitesimal anisotropic Weyl transformation,
\begin{equation}
  \delta_\sigma\tau_\mu=z\sigma\tau_\mu,
  \qquad
  \delta_\sigma h_{\mu\nu}=2\sigma h_{\mu\nu},
  \label{eq:inf-weyl}
\end{equation}
the logarithmic counterterm gives the source-induced one-loop Weyl variation
\begin{equation}
  \delta_\sigma W_{1,\ren}^{\rm sm,(s)}
  =
  \frac{1}{16\pi^2}
  \int\dd^3x\,N\sqrt h\,\sigma\,
  \mathfrak A_{\TTNC,\rm sm}^{(s)} .
  \label{eq:weyl-variation-sector}
\end{equation}
For the homogeneous black brane this gives
\begin{equation}
  z\epsilon_{1,\rm sm}^{(s)}-2p_{1,\rm sm}^{(s)}
  =
  \frac{\calC_1^{(s)}}{16\pi^2}r_h^{z+2}.
  \label{eq:ward-smooth-sector}
\end{equation}
Thus \(\calC_1^{(s)}\) is not only a coefficient in the radial heat-kernel
expansion.  It is the smooth matter contribution to the homogeneous
Lifshitz/TTNC scale Ward identity.

\subsection{Conical entropy and logarithmic thermodynamics}
\label{sec:thermo-ward}

The horizon contribution has a different geometric origin from the smooth
radial logarithm.  It is obtained from the replica variation of the local
effective action on the conical geometry and is localized on the horizon.
It is useful to represent this contribution by the logarithmic free-energy
density
\begin{equation}
  f_{1,\log}^{\Sigma,(s)}
  =
  -\frac{zL^2\calW_h^{(s)}}{16\pi^2}
  r_h^{z+2}\log\frac{r_h}{\mu}.
  \label{eq:flog-horizon-sector}
\end{equation}
Using
\begin{equation}
  T=\frac{z+2}{4\pi}r_h^z,
  \qquad
  \frac{\partial T}{\partial r_h}
  =
  \frac{z(z+2)}{4\pi}r_h^{z-1},
  \label{eq:T-derivative}
\end{equation}
one finds
\begin{equation}
  -\frac{\partial f_{1,\log}^{\Sigma,(s)}}{\partial T}
  =
  \frac{L^2\calW_h^{(s)}}{4\pi}
  r_h^2\log\frac{r_h}{\mu}
  +
  \hbox{non-logarithmic terms},
  \label{eq:horizon-free-energy-entropy-check}
\end{equation}
in agreement with the conical entropy density
\eqref{eq:slog-horizon-general}.

Combining the smooth radial contribution with the horizon-localized
conical contribution gives the logarithmic thermal free-energy density
\begin{equation}
  f_{1,\log}^{\rm therm,(s)}
  =
  -\frac{\calC_{\rm therm}^{(s)}}{16\pi^2}
  r_h^{z+2}\log\frac{r_h}{\mu},
  \qquad
  \calC_{\rm therm}^{(s)}
  =
  \calC_1^{(s)}+zL^2\calW_h^{(s)} .
  \label{eq:flog-therm-sector}
\end{equation}
Consequently,
\begin{equation}
  s_{1,\log}^{\rm therm,(s)}
  =
  \frac{\calC_{\rm therm}^{(s)}}{4\pi z}
  r_h^2\log\frac{r_h}{\mu}
  \label{eq:slog-therm-sector}
\end{equation}
up to scheme-dependent non-logarithmic terms proportional to \(r_h^2\).
In the horizon-only limit
\(\calC_{\rm therm}^{(s)}\to zL^2\calW_h^{(s)}\),
\eqref{eq:slog-therm-sector} reduces to
\eqref{eq:slog-horizon-general}.

The scale dependence of the logarithmic free energies makes the distinction
between the two contributions explicit:
\begin{align}
  \mu\frac{\partial f_{1,\log}^{\rm sm,(s)}}{\partial\mu}
  &=
  \frac{\calC_1^{(s)}}{16\pi^2}r_h^{z+2},
  \label{eq:smooth-scale-sector}
  \\
  \mu\frac{\partial f_{1,\log}^{\Sigma,(s)}}{\partial\mu}
  &=
  \frac{zL^2\calW_h^{(s)}}{16\pi^2}r_h^{z+2},
  \label{eq:horizon-scale-sector}
  \\
  \mu\frac{\partial f_{1,\log}^{\rm therm,(s)}}{\partial\mu}
  &=
  \frac{\calC_{\rm therm}^{(s)}}{16\pi^2}r_h^{z+2}.
  \label{eq:thermal-scale-sector}
\end{align}
For a homogeneous thermal state \(p_1=-f_1\) and
\(\epsilon_1=f_1+Ts_1\).  A purely homogeneous non-logarithmic term
\(f_{\rm hom}=Cr_h^{z+2}\) satisfies
\(z\epsilon_{\rm hom}-2p_{\rm hom}=0\).  The logarithmic smooth term gives
the nonzero source-side Ward projection \eqref{eq:ward-smooth-sector},
whereas the conical term contributes to the logarithmic thermal entropy.
Thus the smooth boundary-source contribution to the homogeneous Ward
combination is
\begin{equation}
  z\epsilon_{\rm sm}^{\rm matter}-2p_{\rm sm}^{\rm matter}
  =
  \frac{\calC_1^{\rm matter}}{16\pi^2}r_h^{z+2}.
  \label{eq:ward-matter}
\end{equation}

The separation between \(\calC_1^{(s)}\) and \(\calW_h^{(s)}\) is therefore
a separation between two local origins of the logarithmic term.  The
coefficient \(\calC_1^{(s)}\) is obtained from the smooth asymptotic radial
integral and fixes the logarithmic counterterm for the boundary TTNC
sources.  The coefficient \(\calW_h^{(s)}\), by contrast, is obtained from
the conical variation of the local heat-kernel action on the replicated
geometry.  It contributes to the logarithmic entropy but is not part of the
smooth boundary-source counterterm.  Equivalently,
\(\calC_{\rm therm}^{(s)}\) governs the logarithmic thermal entropy, while
\(\calC_1^{(s)}\) governs the smooth source-side projection of the
Lifshitz/TTNC Ward identity.

For the fixed probe operator, normalization, and subtraction convention used
here, the displayed logarithmic coefficients are fixed by the local
heat-kernel density.  Finite local counterterms may change
scheme-dependent non-logarithmic terms and, in a general non-flat TTNC
background, may reshuffle Weyl-exact contributions to the anomaly
functional.  They do not change the logarithmic coefficient extracted from
the homogeneous \(dr/r\) term in the smooth radial expansion.  Thus the
smooth TTNC Ward projection is governed by \(\calC_1^{(s)}\), while the
logarithmic thermal entropy is governed by
\(\calC_{\rm therm}^{(s)}=\calC_1^{(s)}+zL^2\calW_h^{(s)}\).

For the Maxwell probe, the determinant is the gauge-fixed Abelian vector
determinant together with the complex Faddeev--Popov ghost determinant
defined in subsection~\ref{subsec:vector-det}.  The corresponding conical
entropy therefore includes the standard spin-one contact contribution at
the horizon~\cite{Kabat:1995eq}.  We use this contact term in the standard gauge-fixed
heat-kernel sense.  It is a contribution to the local conical, or
gravitational, entropy of the Maxwell probe determinant, rather than an
edge-mode-resolved entanglement entropy calculation.  Possible
reorganizations of the Maxwell entropy into bulk and edge contributions are
not addressed here~\cite{Donnelly:2014fua}.  This local coefficient should not be confused with the
full Einstein--Maxwell--dilaton one-loop determinant, in which the metric,
dilaton, and background-supporting gauge fluctuations mix.

\subsection{Source-side heat-kernel representation}
\label{sec:homogeneous-projection}

The previous subsections fixed the homogeneous flat-source projection of the
smooth TTNC logarithm.  The same result has a natural source-side
representation that is useful for extending the calculation to non-flat TTNC
backgrounds.  Let $ \{\tau_\mu,h^{\mu\nu},v^\mu,h_{\mu\nu},M_\mu\}$
denote the boundary TTNC source data, with \(\tau_\mu\) the clock one-form,
\(h^{\mu\nu}\) the inverse spatial metric, \(v^\mu\tau_\mu=-1\), and
\(\tau\wedge\dd\tau=0\) in the twistless sector.  The anisotropic Weyl
transformation acts as
\begin{equation}
  \delta_\sigma\tau_\mu=z\sigma\tau_\mu,
  \qquad
  \delta_\sigma h_{\mu\nu}=2\sigma h_{\mu\nu},
  \qquad
  \delta_\sigma v^\mu=-z\sigma v^\mu,
  \qquad
  \delta_\sigma h^{\mu\nu}=-2\sigma h^{\mu\nu}.
  \label{eq:weyl-ttnc-sources}
\end{equation}
The smooth logarithmic counterterm can be written locally as
\begin{equation}
  \Gamma_{1,\ct}^{\rm sm,(s)}
  =
  -\frac{\log(\Lambda/\mu)}{16\pi^2}
  \int\dd^3x\,e\,
  \mathfrak A_{\TTNC,\rm sm}^{(s)}[\tau,h,v,M],
  \qquad
  e\equiv N\sqrt h .
  \label{eq:local-counterterm-source}
\end{equation}
The corresponding source-induced Weyl variation is
\begin{equation}
  \delta_\sigma W_{1,\ren}^{\rm sm,(s)}
  =
  \frac{1}{16\pi^2}
  \int\dd^3x\,e\,\sigma\,
  \mathfrak A_{\TTNC,\rm sm}^{(s)}[\tau,h,v,M],
  \label{eq:source-weyl-variation}
\end{equation}
or, equivalently,
\begin{equation}
  z\mathcal E_{\rm sm}^{(s)}
  -
  \mathcal T^i{}_{i,\rm sm}^{(s)}
  =
  \frac{1}{16\pi^2}
  \mathfrak A_{\TTNC,\rm sm}^{(s)} .
  \label{eq:source-ward-identity}
\end{equation}

The smooth density is represented by the radial logarithmic coefficient of
the bulk heat-kernel density evaluated on the asymptotically Lifshitz
Fefferman--Graham/TTNC expansion:
\begin{equation}
  \mathfrak A_{\TTNC,\rm sm}^{(s)}
  =
  \left[
  \sqrt g\,a_4^{(s)}[g(\tau,h,v,M)]
  \right]_{\dd r/r}.
  \label{eq:source-heatkernel-rep}
\end{equation}
For general non-flat TTNC sources, this density may be expanded in a basis
of foliation-preserving diffeomorphism invariants of total anisotropic Weyl
weight \(z+2\),
\begin{equation}
  \left[
  \sqrt g\,a_4^{(s)}[g(\tau,h,v,M)]
  \right]_{\dd r/r}
  =
  e\sum_I b_I^{(s)}
  \calI_I^{\TTNC}[\tau,h,v,M].
  \label{eq:ttnc-invariant-expansion}
\end{equation}
The cohomological organization of this invariant basis follows the standard
Lifshitz anomaly analysis
\cite{Baggio:2011ha,Arav:2014goa,Arav:2016xjc,Arav:2016akx,
Auzzi:2015fgg,Auzzi:2016lrq}.  The present calculation fixes the
homogeneous black-brane projection of \eqref{eq:source-heatkernel-rep}.  For
the flat TTNC sources \eqref{eq:flat-ttnc},
\begin{equation}
  \frac{1}{\beta V_2}
  \int\dd^3x\,e\,
  \frac{1}{16\pi^2}
  \mathfrak A_{\TTNC,\rm sm}^{(s)}
  =
  \frac{\calC_1^{(s)}}{16\pi^2}r_h^{z+2},
  \label{eq:homogeneous-source-projection}
\end{equation}
in agreement with \eqref{eq:ward-smooth-sector}.  The horizon coefficient
\(\calW_h^{(s)}\) is separate: it is the replica contribution to the thermal
entropy and enters the total thermal coefficient through
\eqref{eq:sector-coeff-defs}.

\section{The \texorpdfstring{\(z=1\)}{z=1} planar AdS limit}
\label{sec:z1-check}

The relativistic limit \(z=1\) gives the planar Euclidean \(\AdS_4\) black
brane in the coordinates of \eqref{eq:euclidean-lif-bb}.  In this limit
\begin{equation}
  \calC_1^{(\chi)}\big|_{z=1}
  =
  \calC_1^{(\psi)}\big|_{z=1}
  =
  \calC_1^{(A)}\big|_{z=1}
  =
  0 .
  \label{eq:C1-z1-all}
\end{equation}
Therefore the smooth radial logarithm and the smooth flat-boundary
source-induced contribution to \(z\epsilon-2p\) vanish for all probe sectors
considered here.  The remaining logarithmic coefficient is the
horizon-localized conical coefficient.

For the scalar sector,
\begin{equation}
  L^2\calW_h^{(\chi)}\big|_{z=1}
  =
  (m_\chi L)^2\nu_\xi
  -
  12\nu_\xi^2
  +
  \frac{1}{30}.
  \label{eq:z1-scalar-Wh}
\end{equation}
For a massless conformally coupled scalar this gives
\(L^2\calW_h^{(\chi)}=1/30\).  For a massless Dirac spinor and for a Maxwell
vector,
\begin{equation}
  \calC_{\rm therm}^{(\psi)}
  \big|_{z=1,m_\psi=0}
  =
  \frac15,
  \qquad
  \calC_{\rm therm}^{(A)}
  \big|_{z=1}
  =
  \frac25 .
  \label{eq:z1-spin-vector-checks}
\end{equation}
The logarithmic entropy density is
\begin{equation}
  s_{1,\log}^{z=1,(s)}
  =
  \frac{L^2\calW_h^{(s)}}{4\pi}
  r_h^2\log\frac{r_h}{\mu},
  \label{eq:z1-entropy-general}
\end{equation}
which is the local conical heat-kernel result for the planar \(\AdS_4\)
black brane.  Thus the \(z=1\) limit confirms that the smooth coefficient
\(\calC_1\) isolates an intrinsically anisotropic source-side scale response,
while \(\calW_h\) reduces to the standard horizon entropy coefficient.

\section{Discussion}
\label{sec:discussion}

We have computed the logarithmic one-loop matter contribution to a
four-dimensional Lifshitz black brane for scalar, Dirac, and Maxwell probe
fields.  The main physical result is the identification of the matter-induced
logarithmic correction to the homogeneous Lifshitz scale relation.  For flat
TTNC boundary sources, the classical planar black brane satisfies
\(z\epsilon=2p\).  At one loop, the smooth radial logarithm produces a
source-side contribution governed by \(\calC_1\), so that
\eqref{eq:ward-smooth-sector} measures the quantum correction to the Ward
combination \(z\epsilon-2p\).  This correction is distinct from the
horizon-localized conical contribution governed by \(\calW_h\), which enters
the logarithmic entropy.  The total logarithmic thermal coefficient is
therefore
\(\calC_{\rm therm}=\calC_1+zL^2\calW_h\), as in
\eqref{eq:slog-therm-sector}, but the boundary source response is governed
only by \(\calC_1\).

The spin-dependent coefficients make this separation clear.  The scalar
sector shows how the nonminimal coupling \(\nu_\xi\) enters both the smooth
scale response and the conical entropy.  The Dirac sector includes the
fermionic statistics sign, while the Maxwell sector contains the
gauge-fixed vector determinant together with the complex ghost determinant.
In the Maxwell case, the conical coefficient includes the standard spin-one
contact term at the horizon.  These results, summarized in
table~\ref{tab:coefficients}, give the scalar, spinor, and vector
contributions separately and therefore show how spin, statistics, and gauge
constraints affect the source-side Ward response and the horizon entropy in
different ways.

The relativistic planar limit gives a useful physical check on the
interpretation.  At \(z=1\), all smooth radial coefficients vanish, so the
flat-boundary source-induced contribution to \(z\epsilon-2p\) disappears for
the probe sectors considered here.  The remaining logarithmic coefficient is
the local conical heat-kernel entropy coefficient for the planar
\(\AdS_4\) black brane.  For a massless conformally coupled scalar, a
massless Dirac spinor, and a Maxwell vector, the corresponding
\(\calC_{\rm therm}\) values are \(1/30\), \(1/5\), and \(2/5\),
respectively.  Thus the smooth coefficient \(\calC_1\) isolates an
intrinsically anisotropic quantum scale response, rather than a rewriting of
the ordinary relativistic horizon entropy logarithm.

The result also gives a concrete local constraint on the future full
Einstein--Maxwell--dilaton one-loop determinant.  In the complete dynamical
problem, metric, dilaton, and background-supporting gauge fluctuations mix,
and the answer will depend on the corresponding gauge fixing, ghost
determinants, zero modes, and possible operator mixing.  Nevertheless, the
full calculation must reproduce the same distinction between smooth
source-side logarithms and horizon conical logarithms.  The probe scalar,
spinor, and vector coefficients obtained here provide local
building blocks against which the decoupled limits, ghost structure,
spin-one contact term, and \(z=1\) behavior of the full EMD determinant can
be checked.  In this sense the calculation gives nontrivial information
beyond the evaluation of heat-kernel coefficients: it identifies which local
terms correct the Lifshitz Ward identity and which local terms correct the
entropy.

Several extensions follow naturally from this separation.  For non-flat TTNC
sources, one may extract the full invariant expansion in
\eqref{eq:ttnc-invariant-expansion} and determine how the homogeneous
coefficient \(\calC_1\) sits inside the complete source-side anomaly
functional.  Beyond the local logarithmic sector, finite non-local
determinants can be studied by radial determinant methods
\cite{Kirsten:2003py,Gelfand:1959nq}.  These extensions would complement the
present result by adding nonlocal information, but the local distinction
between TTNC scale response and horizon entropy already appears in the
logarithmic sector computed here.

\appendix

\section{Conical heat-kernel coefficients}
\label{app:cone}

The local logarithmic part of a one-loop effective action may be written in the form
\begin{equation}
\Gamma_{1,\log}^{\loc}
=
-\frac{\log(r_h/\mu)}{16\pi^2}
\int_{\calM}\dd^4x\sqrt g\,\mathfrak a_4 .
\label{eq:app-Gamma-log}
\end{equation}
Only the dependence of this local term on the conical opening angle is needed for the logarithmic entropy.  Let \(\calM_n\) denote the Euclidean replicated geometry with opening angle \(2\pi n\) around the horizon two-surface \(\Sigma\).  For a stationary Killing horizon the extrinsic curvatures of \(\Sigma\) vanish, and the conical-defect formulas give, to first order in \(1-n\),
\begin{align}
\int_{\calM_n}\sqrt g\,\calR
&=
n\int_{\calM_1}\sqrt g\,\calR
+4\pi(1-n)\int_\Sigma\sqrt\gamma ,
\label{eq:cone-R}\\
\int_{\calM_n}\sqrt g\,\calR^2
&=
n\int_{\calM_1}\sqrt g\,\calR^2
+8\pi(1-n)\int_\Sigma\sqrt\gamma\,\calR ,
\label{eq:cone-R2}\\
\int_{\calM_n}\sqrt g\,\calR_{MN}\calR^{MN}
&=
n\int_{\calM_1}\sqrt g\,\calR_{MN}\calR^{MN}
+4\pi(1-n)\int_\Sigma\sqrt\gamma\,\calR_\perp ,
\label{eq:cone-Ricci2}\\
\int_{\calM_n}\sqrt g\,\calR_{MNRS}\calR^{MNRS}
&=
n\int_{\calM_1}\sqrt g\,\calR_{MNRS}\calR^{MNRS}
+8\pi(1-n)\int_\Sigma\sqrt\gamma\,\calR_{\perp\perp} .
\label{eq:cone-Riemann2}
\end{align}
For a density of the form \eqref{eq:generic-a4}, these relations imply
\begin{equation}
\int_{\calM_n}\sqrt g\,\mathfrak a_4
=
n\int_{\calM_1}\sqrt g\,\mathfrak a_4
+4\pi(1-n)\int_\Sigma\sqrt\gamma\,\calW_h
+O((1-n)^2),
\label{eq:app-a4-cone}
\end{equation}
with \(\calW_h\) given by \eqref{eq:generic-Wh}.  The entropy is obtained from
\begin{equation}
S_{1,\log}
=
\left.
\left(n\frac{\partial}{\partial n}-1\right)
\Gamma_{1,\log}^{\loc}[\calM_n]
\right|_{n=1}
=
\frac{\log(r_h/\mu)}{4\pi}
\int_\Sigma\sqrt\gamma\,\calW_h .
\label{eq:app-replica-entropy}
\end{equation}
For the planar Lifshitz black brane,
\begin{equation}
\frac{1}{V_2}\int_\Sigma\sqrt\gamma=L^2r_h^2,
\label{eq:app-horizon-area}
\end{equation}
and therefore \eqref{eq:slog-horizon-general} follows.

\section{Curvature expansion and spin-dependent coefficients}
\label{app:curvature-expansion}

The smooth radial logarithm is determined by the term linear in \(u=(r_h/r)^{z+2}\) in the large-\(r\) expansion of \(L^4\mathfrak a_4\).  The required curvature expansions are
\begin{align}
L^2\calR&=R_0+R_1u+O(u^2),
& R_1&=-2(z-1),
\label{eq:app-R1}\\
L^4\calR_{MNRS}\calR^{MNRS}&=K_0+K_1u+O(u^2),
& K_1&=8(z-1)^3,
\label{eq:app-K1}\\
L^4\calR_{MN}\calR^{MN}&=P_0+P_1u+O(u^2),
& P_1&=4(z-1)(z^2+2),
\label{eq:app-P1}\\
L^4\calR^2&=S_0+S_1u+O(u^2),
& S_1&=8(z-1)(z^2+2z+3).
\label{eq:app-S1}
\end{align}
If
\begin{equation}
\mathfrak a_4
=
\alpha\calR_{MNRS}\calR^{MNRS}
+\beta\calR_{MN}\calR^{MN}
+\gamma\calR^2
+\delta m^2\calR+\text{terms without curvature},
\label{eq:app-generic-density}
\end{equation}
then the smooth coefficient is
\begin{equation}
\calC_1=\alpha K_1+\beta P_1+\gamma S_1+\delta(mL)^2R_1 .
\label{eq:app-C1-master}
\end{equation}
For the scalar, \((\alpha,\beta,\gamma,\delta)=(1/180,-1/180,\nu_\xi^2/2,\nu_\xi)\), which gives \eqref{eq:C1-scalar}.  For the Dirac spinor, \((\alpha,\beta,\gamma,\delta)=(7/360,1/45,-1/72,-1/3)\), which gives \eqref{eq:C1-spinor}.  For the Maxwell gauge-plus-ghost sector, \((\alpha,\beta,\gamma)=(-13/180,88/180,-25/180)\), which gives \eqref{eq:C1-vector}.

Combining \eqref{eq:app-C1-master} with the conical formula \eqref{eq:generic-Wh} yields the coefficients in table~\ref{tab:coefficients}.  In each case the common factor \(z-1\) in \(\calC_1\) gives the absence of a smooth radial logarithm in the relativistic planar \(\AdS_4\) limit.

\section*{Conflict of interest}
The authors declare no competing interests.

\section*{Data availability statement}
No new data were created or analysed in this study.

\section*{Ethics statement}
This work did not involve human participants, human data, human tissue, animals, or any activity requiring ethics approval.

\funding{The authors received no specific funding for this work.}

\bibliographystyle{JHEP}
\bibliography{reference}

@article{Kabat:1995eq,
    author = "Kabat, Daniel N.",
    title = "{Black hole entropy and entropy of entanglement}",
    eprint = "hep-th/9503016",
    archivePrefix = "arXiv",
    reportNumber = "RU-95-06",
    doi = "10.1016/0550-3213(95)00443-V",
    journal = "Nucl. Phys. B",
    volume = "453",
    pages = "281--299",
    year = "1995"
}

@article{Donnelly:2014fua,
    author = "Donnelly, William and Wall, Aron C.",
    title = "{Entanglement entropy of electromagnetic edge modes}",
    eprint = "1412.1895",
    archivePrefix = "arXiv",
    primaryClass = "hep-th",
    doi = "10.1103/PhysRevLett.114.111603",
    journal = "Phys. Rev. Lett.",
    volume = "114",
    number = "11",
    pages = "111603",
    year = "2015"
}

@article{Tarrio:2011de,
   title={Black holes and black branes in Lifshitz spacetimes},
   volume={2011},
   ISSN={1029-8479},
   url={http://dx.doi.org/10.1007/JHEP09(2011)017},
   DOI={10.1007/jhep09(2011)017},
   number={9},
   journal={Journal of High Energy Physics},
   publisher={Springer Science and Business Media LLC},
   author={Tarrío, Javier and Vandoren, Stefan},
   year={2011},
   month=Sept }

@article{Quinta:2016cvy,
   title={Vacuum polarization in asymptotically Lifshitz black holes},
   volume={93},
   ISSN={2470-0029},
   url={http://dx.doi.org/10.1103/PhysRevD.93.124073},
   DOI={10.1103/physrevd.93.124073},
   number={12},
   journal={Physical Review D},
   publisher={American Physical Society (APS)},
   author={Quinta, Gonçalo M. and Flachi, Antonino and Lemos, José P. S.},
   year={2016},
   month=June }

@article{Christensen:2013rfa,
   title={Torsional Newton-Cartan geometry and Lifshitz holography},
   volume={89},
   ISSN={1550-2368},
   url={http://dx.doi.org/10.1103/PhysRevD.89.061901},
   DOI={10.1103/physrevd.89.061901},
   number={6},
   journal={Physical Review D},
   publisher={American Physical Society (APS)},
   author={Christensen, Morten H. and Hartong, Jelle and Obers, Niels A. and Rollier, B.},
   year={2014},
   month=Mar }

@article{Christensen:2013lma,
   title={Boundary stress-energy tensor and Newton-Cartan geometry in Lifshitz holography},
   volume={2014},
   ISSN={1029-8479},
   url={http://dx.doi.org/10.1007/JHEP01(2014)057},
   DOI={10.1007/jhep01(2014)057},
   number={1},
   journal={Journal of High Energy Physics},
   publisher={Springer Science and Business Media LLC},
   author={Christensen, Morten H. and Hartong, Jelle and Obers, Niels A. and Rollier, Blaise},
   year={2014},
   month=Jan }

@article{Baggio:2011ha,
  title={Anomalous breaking of anisotropic scaling symmetry in the quantum lifshitz model},
   volume={2012},
   ISSN={1029-8479},
   url={http://dx.doi.org/10.1007/JHEP07(2012)099},
   DOI={10.1007/jhep07(2012)099},
   number={7},
   journal={Journal of High Energy Physics},
   publisher={Springer Science and Business Media LLC},
   author={Baggio, Marco and de Boer, Jan and Holsheimer, Kristian},
   year={2012},
   month=July }

@article{Hartong:2022lsy,
   title={Lifshitz hydrodynamics from Lifshitz black branes with linear momentum},
   volume={2016},
   ISSN={1029-8479},
   url={http://dx.doi.org/10.1007/JHEP10(2016)120},
   DOI={10.1007/jhep10(2016)120},
   number={10},
   journal={Journal of High Energy Physics},
   publisher={Springer Science and Business Media LLC},
   author={Hartong, Jelle and Obers, Niels A. and Sanchioni, Marco},
   year={2016},
   month=Oct }

@article{Vassilevich:2003xt,
   title={Heat kernel expansion: user’s manual},
   volume={388},
   ISSN={0370-1573},
   url={http://dx.doi.org/10.1016/j.physrep.2003.09.002},
   DOI={10.1016/j.physrep.2003.09.002},
   number={5-6},
   journal={Physics Reports},
   publisher={Elsevier BV},
   author={Vassilevich, D.V.},
   year={2003},
   month=Dec, pages={279–360} }

@article{Fursaev:1995ef,
  title={Description of the Riemannian geometry in the presence of conical defects},
   volume={52},
   ISSN={0556-2821},
   url={http://dx.doi.org/10.1103/PhysRevD.52.2133},
   DOI={10.1103/physrevd.52.2133},
   number={4},
   journal={Physical Review D},
   publisher={American Physical Society (APS)},
   author={Fursaev, Dmitri V. and Solodukhin, Sergey N.},
   year={1995},
   month=Aug, pages={2133–2143} }

@article{Solodukhin:2011gn,
   title={Entanglement Entropy of Black Holes},
   volume={14},
   ISSN={1433-8351},
   url={http://dx.doi.org/10.12942/lrr-2011-8},
   DOI={10.12942/lrr-2011-8},
   number={1},
   journal={Living Reviews in Relativity},
   publisher={Springer Science and Business Media LLC},
   author={Solodukhin, Sergey N.},
   year={2011},
   month=Oct }

@article{Kachru:2008yh,
   title={Gravity duals of Lifshitz-like fixed points},
   volume={78},
   ISSN={1550-2368},
   url={http://dx.doi.org/10.1103/PhysRevD.78.106005},
   DOI={10.1103/physrevd.78.106005},
   number={10},
   journal={Physical Review D},
   publisher={American Physical Society (APS)},
   author={Kachru, Shamit and Liu, Xiao and Mulligan, Michael},
   year={2008},
   month=Nov }

@article{Taylor:2008tg,
      title={Non-relativistic holography},
      author={Marika Taylor},
      year={2008},
      eprint={0812.0530},
      archivePrefix={arXiv},
      primaryClass={hep-th},
      url={https://arxiv.org/abs/0812.0530},
}

@article{Taylor:2015glc,
   title={Lifshitz holography},
   volume={33},
   ISSN={1361-6382},
   url={http://dx.doi.org/10.1088/0264-9381/33/3/033001},
   DOI={10.1088/0264-9381/33/3/033001},
   number={3},
   journal={Classical and Quantum Gravity},
   publisher={IOP Publishing},
   author={Taylor, Marika},
   year={2016},
   month=Jan, pages={033001} }

@article{Bertoldi:2009vn,
   title={Black holes in asymptotically Lifshitz spacetimes with arbitrary critical exponent},
   volume={80},
   ISSN={1550-2368},
   url={http://dx.doi.org/10.1103/PhysRevD.80.126003},
   DOI={10.1103/physrevd.80.126003},
   number={12},
   journal={Physical Review D},
   publisher={American Physical Society (APS)},
   author={Bertoldi, Gaetano and Burrington, Benjamin A. and Peet, Amanda},
   year={2009},
   month=Dec }

@article{Bertoldi:2009dt,
   title={Thermodynamics of black branes in asymptotically Lifshitz spacetimes},
   volume={80},
   ISSN={1550-2368},
   url={http://dx.doi.org/10.1103/PhysRevD.80.126004},
   DOI={10.1103/physrevd.80.126004},
   number={12},
   journal={Physical Review D},
   publisher={American Physical Society (APS)},
   author={Bertoldi, Gaetano and Burrington, Benjamin A. and Peet, Amanda W.},
   year={2009},
   month=Dec }

@article{Mann:2009yx,
      title={Lifshitz Topological Black Holes},
      author={R. B. Mann},
      year={2009},
      eprint={0905.1136},
      archivePrefix={arXiv},
      primaryClass={hep-th},
      url={https://arxiv.org/abs/0905.1136},
}

@article{Ross:2011gu,
   title={Holography for asymptotically locally Lifshitz spacetimes},
   volume={28},
   ISSN={1361-6382},
   url={http://dx.doi.org/10.1088/0264-9381/28/21/215019},
   DOI={10.1088/0264-9381/28/21/215019},
   number={21},
   journal={Classical and Quantum Gravity},
   publisher={IOP Publishing},
   author={Ross, Simon F},
   year={2011},
   month=Oct, pages={215019} }

@article{Bergshoeff:2014uea,
      title={Torsional Newton-Cartan Geometry and the Schr\"odinger Algebra},
      author={Eric A. Bergshoeff and Jelle Hartong and Jan Rosseel},
      year={2014},
      eprint={1409.5555},
      archivePrefix={arXiv},
      primaryClass={hep-th},
      url={https://arxiv.org/abs/1409.5555},
}

@article{Hartong:2015zia,
   title={Hořava-Lifshitz gravity from dynamical Newton-Cartan geometry},
   volume={2015},
   ISSN={1029-8479},
   url={http://dx.doi.org/10.1007/JHEP07(2015)155},
   DOI={10.1007/jhep07(2015)155},
   number={7},
   journal={Journal of High Energy Physics},
   publisher={Springer Science and Business Media LLC},
   author={Hartong, Jelle and Obers, Niels A.},
   year={2015},
   month=July }

@article{Arav:2014goa,
   title={Lifshitz scale anomalies},
   volume={2015},
   ISSN={1029-8479},
   url={http://dx.doi.org/10.1007/JHEP02(2015)078},
   DOI={10.1007/jhep02(2015)078},
   number={2},
   journal={Journal of High Energy Physics},
   publisher={Springer Science and Business Media LLC},
   author={Arav, Igal and Chapman, Shira and Oz, Yaron},
   year={2015},
   month=Feb }

@article{Arav:2016xjc,
   title={Non-relativistic scale anomalies},
   volume={2016},
   ISSN={1029-8479},
   url={http://dx.doi.org/10.1007/JHEP06(2016)158},
   DOI={10.1007/jhep06(2016)158},
   number={6},
   journal={Journal of High Energy Physics},
   publisher={Springer Science and Business Media LLC},
   author={Arav, Igal and Chapman, Shira and Oz, Yaron},
   year={2016},
   month=June }

@article{Arav:2016akx,
   title={Lifshitz anomalies, Ward identities and split dimensional regularization},
   volume={2017},
   ISSN={1029-8479},
   url={http://dx.doi.org/10.1007/JHEP03(2017)088},
   DOI={10.1007/jhep03(2017)088},
   number={3},
   journal={Journal of High Energy Physics},
   publisher={Springer Science and Business Media LLC},
   author={Arav, Igal and Oz, Yaron and Raviv-Moshe, Avia},
   year={2017},
   month=Mar }

@article{Auzzi:2015fgg,
    author = "Auzzi, Roberto and Baiguera, Stefano and Nardelli, Giuseppe",
    title = "{On Newton-Cartan trace anomalies}",
    eprint = "1511.08150",
    archivePrefix = "arXiv",
    primaryClass = "hep-th",
    doi = "10.1007/JHEP02(2016)177",
    journal = "JHEP",
    volume = "02",
    pages = "003",
    year = "2016",
    note = "[Erratum: JHEP 02, 177 (2016)]"
}

@article{Auzzi:2016lrq,
   title={Heat kernel for Newton-Cartan trace anomalies},
   volume={2016},
   ISSN={1029-8479},
   url={http://dx.doi.org/10.1007/JHEP07(2016)047},
   DOI={10.1007/jhep07(2016)047},
   number={7},
   journal={Journal of High Energy Physics},
   publisher={Springer Science and Business Media LLC},
   author={Auzzi, Roberto and Nardelli, Giuseppe},
   year={2016},
   month=July }

@article{Cong:2024lfh,
      title={Holographic dictionary for Lifshitz and hyperscaling violating black holes},
      author={Wan Cong and David Kubizňák and Robert B. Mann and Manus R. Visser},
      year={2025},
      eprint={2410.16145},
      archivePrefix={arXiv},
      primaryClass={hep-th},
      url={https://arxiv.org/abs/2410.16145},
}

@article{Kirsten:2003py,
    author = "Kirsten, Klaus and McKane, Alan J.",
    title = "{Functional determinants by contour integration methods}",
    eprint = "math-ph/0305010",
    archivePrefix = "arXiv",
    doi = "10.1016/S0003-4916(03)00149-0",
    journal = "Annals Phys.",
    volume = "308",
    pages = "502--527",
    year = "2003"
}

@article{Gelfand:1959nq,
    author = "Gelfand, I. M. and Yaglom, A. M.",
    title = "{Integration in functional spaces and it applications in quantum physics}",
    doi = "10.1063/1.1703636",
    journal = "J. Math. Phys.",
    volume = "1",
    pages = "48",
    year = "1960"
}

\end{document}